\documentclass[lettersize,journal]{IEEEtran}
\usepackage{amsmath,amsfonts}
\usepackage{algorithmic}
\usepackage{algorithm}
\usepackage{array}
\usepackage[caption=false,font=normalsize,labelfont=sf,textfont=sf]{subfig}
\usepackage{textcomp}
\usepackage{stfloats}
\usepackage{url}
\usepackage{verbatim}
\usepackage{graphicx}
\usepackage{cite}
\usepackage{multirow}
\usepackage{diagbox}
\usepackage{booktabs}

\usepackage[colorlinks  = true,
            linkcolor   = black,
            urlcolor    = magenta,
            anchorcolor = blue,
            bookmarks   = false
            ]{hyperref}
\begin{document}

\title{Data Augmentation of Bridging the Delay Gap for DL-based Massive MIMO CSI Feedback}

\author{Hengyu Zhang, Zhilin Lu, Xudong Zhang, and Jintao Wang,~\IEEEmembership{Senior Member,~IEEE}
\thanks{The authors are with the Department of Electronic Engineering, Tsinghua
University, Beijing 100084, China, and also with the Beijing National Research Center for Information Science and Technology (BNRist), Tsinghua University, Beijing 100084, China (e-mail: zhanghen23@mails.tsinghua.edu.cn; luzl18@mails.tsinghua.edu.cn; zxd22@mails.tsinghua.edu.cn; jintaowang@tsinghua.edu.cn). \textit{(Corresponding author: Jintao Wang.)}}
}



\maketitle

\begin{abstract}
In massive multiple-input multiple-output (MIMO) systems under the frequency division duplexing (FDD) mode, the user equipment (UE) needs to feed channel state information (CSI) back to the base station (BS). Though deep learning approaches have made a hit in the CSI feedback problem, whether they can remain excellent in actual environments needs to be further investigated. In this letter, we point out that the real-time dataset in application often has the domain gap from the training dataset caused by the time delay. To bridge the gap, we propose bubble-shift (B-S) data augmentation, which attempts to offset performance degradation by changing the delay and remaining the channel information as much as possible. Moreover, random-generation (R-G) data augmentation is especially proposed for outdoor scenarios due to the complex distribution of its channels. It generalizes the characteristics of the channel matrix and alleviates the over-fitting problem. Simulation results show that the proposed data augmentation boosts the robustness of networks in both indoor and outdoor environments. The open source codes are available at \href{https://github.com/zhanghy23/CRNet-Aug}{https://github.com/zhanghy23/CRNet-Aug}.
\end{abstract}


\section{Introduction}
\IEEEPARstart{T}{he} massive multiple-input multiple-output (MIMO) system is crucial to 5G communication systems\cite{chen2016measurement}\cite{vieira2014flexible}. Equipped with a large number of antennas, the base station (BS) requires real-time channel state information (CSI) to conduct beamforming for performance gain. In massive MIMO systems under the frequency division duplexing (FDD) mode, the user equipment (UE) needs to detect and feed CSI back to BS, since uplink and downlink channels share no reciprocity. However, a huge number of antennas result in enormous overhead for direct transmission of CSI. Therefore, the CSI matrix must be compressed before being transmitted. The traditional methods based on compressed sensing (CS) exploit the sparsity of the CSI matrix\cite{9695409}. However, the practical system can not completely satisfy this requirement with the large ratio compression.

Deep learning (DL) approaches have been widely utilized to compress the CSI matrix in recent years. CsiNet\cite{wen2018deep} firstly employ the encoder-decoder network structure, setting followers a benchmark. In \cite{lu2020multi}, CRNet adopts multi-resolution convolution kernels to extract features since the sparsity of CSI varies in different scenes. CLNet introduced in \cite{ji2021clnet} achieves excellent performance based on the attention mechanism with less computational overhead. 

To enhance DL approaches training for CSI feedback, some research regard the CSI matrix as pictures to perform data augmentation. Liu \cite{liu2022training} develops a model-driven augmentation approach by cyclically shifting the magnitude and selecting
the uniform distribution as the augmented phase distribution. Ji \cite{ji2022enhancing} proposes a jigsaw puzzles aided training strategy (JPTS), which aims to maximize the mutual information in local regions by predicting the proper locations of puzzle fragments. Xiao \textit{et al.} \cite{xiao2022ai} exploits noise injection, flipping, shift, and rotation enhancement to alleviate the over-fitting problem.

Commonly, there exists a domain gap between the environment where we train the network and scenarios in real applications. However, most of the existing works and data augmentation approaches are tested in ideal environments, which means that the training dataset and testing dataset are generated in the same scenario. Faced up with the domain gap between the training dataset and testing dataset, whether these networks can keep excellent performance and whether data augmentation approaches above can bridge the gap remain to be verified.

In this letter, we simulate two possible cases in reality to test existing networks and data augmentation approaches. Specifically, the training dataset and testing dataset are generated under different UE moving modes or ranges in our cases. Simulation results demonstrate a decline in the compression accuracy of existing networks compared to the performance when their training and testing scenarios are the same. Certain data augmentation methods proposed in the literature exhibit limited effectiveness in these scenarios, as they primarily focus on enhancing the learning of features present in the training dataset. To address the domain gap resulting from diverse UE motion modes or ranges, we introduce the bubble-shift (B-S) data augmentation technique. Besides, we acknowledge the complexity of outdoor scenes and propose the random-generation (R-G) data augmentation specifically tailored for outdoor scenes. By leveraging our proposed data augmentation techniques, the performance of existing networks is significantly improved.

The main contributions of this letter are listed as follows.

\begin{itemize}
\item{B-S data augmentation is proposed to make up for the domain gap in terms of the time delay. It circularly shifts the CSI matrix in the time delay domain and ensures the integrity of channel characteristics through bubble sort.}
\item{R-G data augmentation is designed for outdoor scenes, which helps mitigate the problem of over-fitting by generating possible channel conditions randomly.}
\item{Scenarios closer to reality are considered in CSI feedback, where there is a domain gap between the testing dataset and the training dataset. The open source datasets are available at https://github.com/zhanghy23/CRNet-Aug.}
\end{itemize}

\section{System Model}
We consider a single-cell massive MIMO system of FDD mode with ${N_t}$ antennas at the BS and ${N_r}$ antennas at the UE. For simplicity, we take ${N_r} = 1$. The number of orthogonal frequency division multiplexing (OFDM) sub-carriers is ${N_c}$. We can describe the received signal ${\bf{y}} \in {\mathbb{C}^{{N_c} \times 1}}$ as follows:
\begin{equation}\label{eq1}
{\bf{y = Ax + z}},
\end{equation}
where ${\bf{x}}$, ${\bf{z}} \in {\mathbb{C}^{{N_c} \times 1}}$ indicate the transmitted signal and additive Gaussian noise respectively. In addition, ${\bf{A}} = {\rm{diag}}({\bf{h}}_1^H{{\bf{p}}_1}, \cdot  \cdot  \cdot ,{\bf{h}}_n^H{{\bf{p}}_n})$ represents equivalent channel matrix where ${{\bf{h}}_i} \in {\mathbb{C}^{{N_t} \times 1}}$ indicates the downlink channel response vector and ${{\bf{p}}_i} \in {\mathbb{C}^{{N_t} \times 1}}$, $i \in \{ 1, \cdot  \cdot  \cdot ,{N_c}\}$ is the beamforming vector at sub-carrier ${i}$.

The BS needs downlink channel matrix ${\bf{H}} = {[{{\bf{h}}_1} \cdot  \cdot  \cdot {{\bf{h}}_{{N_c}}}]^H}$ to decide its beamforming vectors. Due to the non-reciprocity between uplink and downlink channels, the channel matrix ${\bf{H}}$ with ${{2N_c}{N_t}}$ elements needs to be detected and fed back from the UE. Direct transmission leads to unacceptable overhead for the system. Therefore, DL approaches have been used to compress the CSI matrix ${\bf{H}}$.

Most methods utilize the Fourier transform, transferring ${\bf{H}}$ in the spatial-frequency domain to ${{\bf{H}}^{'}}$ in the angular-delay domain since ${{\bf{H}}^{'}}$ is sparse.

\begin{equation}
{\bf{H}}^{'} = {\bf{F}}_c{\bf{HF}}_t^H \label{eq2},
\end{equation}
where ${{\bf{F}}_c} \in {N_c} \times {N_c}$ and ${{\bf{F}}_t} \in {N_t} \times {N_t}$ are both discrete Fourier transform (DFT) matrices. Since the delay in multipath is in a limited range, the first ${N_a}$ rows of ${{\bf{H}}^{'}}$ have large enough values. The elements left are near zero values. We finally truncate the first ${N_a}$ rows of the matrix ${{\bf{H}}^{'}}$ and denote it as ${{\bf{H}}_a}$.

In\cite{wen2018deep,lu2020multi,ji2021clnet}, the proposed networks all follow the structure of auto-encoder\cite{hinton2006reducing}. The UE uses the encoder to compress ${{\bf{H}}_a}$ into a short feature vector ${\bf{v}}$ and only transfers short ${\bf{v}}$ to the BS. The BS exploits the decoder to recover it into ${{\hat{\bf{H}}}_a}$. The whole process can be described below.

\begin{equation}
{{\hat{\bf{H}}}_a}=\mathcal{D}(\mathcal{E}({{\bf{H}}_a},\Theta _\mathcal{E}),\Theta _{\mathcal{D}})\label{eq3},
\end{equation}
where ${\mathcal{D}}$ and ${\mathcal{E}}$ represent the function of encoder and decoder respectively. $\Theta _{\mathcal{D}}$ and $\Theta _\mathcal{E}$ denote corresponding network parameters. The network selects the proper loss function to perform the gradient descent, aiming to minimize the distance between ${{\bf{H}}_a}$ and ${{\hat{\bf{H}}}_a}$.

\section{Proposed Data Augmentation Approaches}
\subsection{Domain Gap in Practical Environments}
Serious domain gap exists in the practical environment between the real-time dataset (testing dataset) and the training dataset. Fig. 1 shows two possible cases in reality. Fig. 1 (a) demonstrates the scenario where the users' motion mode changes. The BS is at the center of the area. Users in the training dataset move on the green fixed path. However, UEs of the testing dataset can move at any blue path, which means they can not only move like UEs of the training dataset but also have more diverse motion modes. Since different motion modes may lead to different channel characteristics, this case can pose a challenge to the network.

Another scenario where users' motion range changes is shown in Fig. 1 (b). Compared with the training dataset in the green area, the UEs in the testing dataset can appear and move in a broader blue area randomly. In this case, networks can only learn the CSI matrix from samples in the training dataset but are short of information of the different CSI in the real employment scenarios.

\subsection{Bubble-Shift Data Augmentation}
Since the domain gap between the training dataset and the testing dataset can cause a decline in performance for the networks, we design the data augmentation approach to bridge the gap. For ${{\bf{H}}_a}$ in the angular-delay domain, information related to time delay has a clear physical meaning. Assuming the domain gap is caused by time delay, we propose B-S data augmentation to improve the robustness of networks under varying time delay. 

\begin{figure}[!t]
\centering
\includegraphics[width=\linewidth]{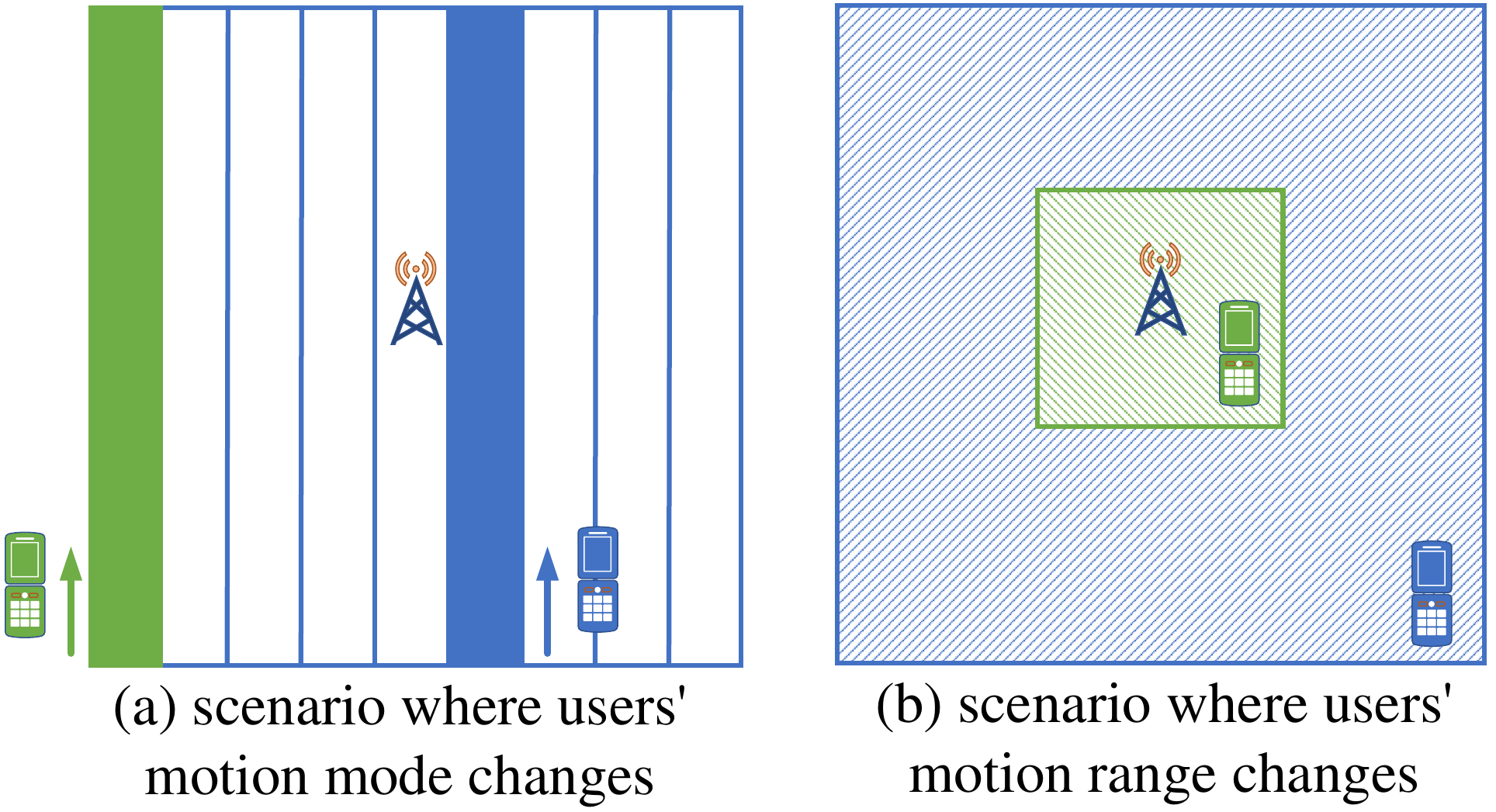}
\caption{Two possible scenarios in reality. The area in green represents the users' motion range of the training dataset and users of the testing dataset appear and move in the blue area.}
\label{fig_1}
\end{figure}

In ${{\bf{H}}_a}$ matrix, if elements with huge values are in the first several rows, it means a short time delay. Oppositely, the time delay can be longer if the first several rows are near zero. Therefore, the B-S approach is to change the internal structure of the matrix and shorten or lengthen the delay in the training dataset, keeping as many original channel features as possible.

The ${{\bf{H}}_a}$ matrix is composed of complex numbers with the corresponding channel information. Therefore, we first divide the ${{\bf{H}}_a}$ matrix to the amplitude and phase matrix. The phase matrix shows a certain continuity and circularity instead of complete randomness. Meanwhile, the phase has no effect on the time delay. As a result, we decide to remain the phase matrix unchanged to reserve the channel information in the training dataset and only operate on the amplitude matrix. 

We classify our algorithm into B-S (Upward) and B-S (Downward), which deal with the amplitude of ${{\bf{H}}_a}$ matrix to change the delay. B-S (Upward) can shorten the time delay of the training dataset shown in Algorithm 1 and B-S (Downward) can lengthen the time delay demonstrated in Algorithm 2.

In Algorithm 1, ${{\bf{H}}_a}\in {N_a} \times {N_t}$ and ${{\bf{H}}_A}\in {N_a} \times {N_t}$ indicate amplitudes of CSI matrix before and after augmentation respectively. ${S} \in \mathbb{C}$ is the hyper-parameter which represents the cyclic shifting step. Each column of ${{\bf{H}}_a}$ is transferred separately. We circularly shift each column upward to shorten the delay. After every cyclic shifting, we should bubble the bottom or top element and shift its position to be consistent with the channel characteristics. Fig. 2 demonstrates the B-S of a one-step cyclic shifting upward. Due to the influence of DFT, we find the value on both sides of the maximum amplitude decrease in turn generally. After we make the cyclic shifting for one step, the top element will reach the bottom as Fig. 2 shows. To arrange the most proper position for the bottom element, we compare it with the element above to bubble gradually. Finally, we reset its position and conduct the next cyclic shifting. B-S (Downward) is similar, which can be acquired in Algorithm 2.\par
\begin{algorithm}[t]
\caption{B-S (Upward)}\label{alg:alg1}
\begin{algorithmic}
\STATE\hspace{-0.3cm}${\textbf{function } \text{B-S} \text{ (} {{\bf{H}}_a} \text{, }\mathit{S}\text{)}}$\\
\STATE\hspace{0.2cm}$\textbf{for }{i = 1,...,N_t}$\\
\STATE \hspace{0.6cm}$step \gets 1$\\
\STATE \hspace{0.6cm}$M \gets argmax \{ {{\bf{H}}_a}(:,i)\}$ \\
\STATE \hspace{0.6cm}$ \textbf{while } step \le min \{ S,M \}$\\
\STATE \hspace{1.1cm}$ \textit{circularly shift } {{\bf{H}}_a}(:,i) \textit{ upward for one step}$\\
\STATE \hspace{1.1cm}$ K \gets N_a$\\
\STATE \hspace{1.1cm}$ \textbf{while } {{\bf{H}}_a}(K,i)>{{\bf{H}}_a}(K-1,i)$\\
\STATE \hspace{1.6cm}$ \textit{exchange } {{\bf{H}}_a}(K,i) \textit{ and } {{\bf{H}}_a}(K-1,i)$\\
\STATE \hspace{1.6cm}$ K \gets K-1$\\
\STATE \hspace{1.1cm}$ \textbf{end while}$\\
\STATE \hspace{1.1cm}$ step \gets step+1$\\
\STATE \hspace{0.6cm}$ \textbf{end while}$\\
\STATE \hspace{0.2cm}$ \textbf{end for}$\\
\STATE \hspace{0.2cm}$ {{\bf{H}}_A} \gets {{\bf{H}}_a} $
\end{algorithmic}
\label{alg1}
\end{algorithm}

In the B-S algorithm, there is no new element generated. We only change the delay and remain the channel information as much as possible. It is worth mentioning that the decision between B-S (Upward) and B-S (Downward) depends on the actual situation. If the feedback precision of the CSI matrix with a short time delay is poor, we can apply B-S (Upward) to enhance the training dataset. Correspondingly, B-S (Downward) can be employed to deal with the situation where networks cannot recover the CSI matrix with a long time delay accurately.

\subsection{Random-Generation Data Augmentation}
Since outdoor environments are more complex and challenging, it is more difficult for the network to extract enough features in the CSI matrix. We design the R-G data augmentation to enhance the channel characteristics. Fig. 3 describes the procedure of R-G. Similarly, the algorithm only focuses on the amplitude of the CSI matrix and the phase remains the same.

As shown in Fig. 3, the matrix picture in outdoor scenes is colorful and rich of the channel information. Most of the highlights appear in blocks due to the geographic adjacency. However, it is difficult to summarize the law of specific values and locations for these highlighted blocks due to the complex outdoor channel environment. The training dataset cannot include all the possible locations and values of these highlighted blocks, in which case, networks may fall into the trap of over-fitting. The goal of R-G is to generate more possible highlighted blocks in possible locations, enabling networks to extract more general features and mitigate the over-fitting problem.     

In the schematic of R-G, we first select the row on which the maximum of the CSI matrix is located and randomly select one element on this row. Then, we generate a $k \times k$ square block centered on this element ($k = 4$ in Fig. 3). Each element in this square block is generated randomly between the maximum and the minimum in uniform distribution. In this case, the time delay of the CSI matrix is not destroyed and more possible channel features are added to the CSI matrix.
\section{Simulation Results and Analysis}
\subsection{Experiment Setting}
We use COST2100\cite{liu2012cost} to generate the data in the indoor environment at 5.3GHz and outdoor environment at 300MHz. The two cases described in Fig. 1 are further concretized.

In the scenario where users' motion mode changes, we set UEs in the training dataset to move at the farthest vertical path away from the BS. UEs in the testing dataset can move at any vertical path. The indoor environment has a square area with a 20m length and UEs move on the vertical path at a speed of 0.1m/s. The outdoor environment shares a square length of 400m and UE's speed is 2m/s.

\begin{algorithm}[t]
\caption{B-S (Downward)}\label{alg:alg2}
\begin{algorithmic}
\STATE\hspace{-0.3cm}${\textbf{function } \text{B-S} \text{ (} {{\bf{H}}_a} \text{, }\mathit{S}\text{)}}$\\
\STATE\hspace{0.2cm}$\textbf{for }{i = 1,...,N_t}$\\
\STATE \hspace{0.6cm}$step \gets 1$\\
\STATE \hspace{0.6cm}$M \gets argmax \{ {{\bf{H}}_a}(:,i)\}$ \\
\STATE \hspace{0.6cm}$ \textbf{while } step \le min \{ S,{N_a}-M \}$\\
\STATE \hspace{1.1cm}$ \textit{circularly shift } {{\bf{H}}_a}(:,i) \textit{ downward for one step}$\\
\STATE \hspace{1.1cm}$ K \gets N_a$\\
\STATE \hspace{1.1cm}$ \textbf{while } {{\bf{H}}_a}(1,i)<{{\bf{H}}_a}(K,i)<{{\bf{H}}_a}(2,i)$\\
\STATE \hspace{1.6cm}$ \textit{exchange } {{\bf{H}}_a}(K,i) \textit{ and } {{\bf{H}}_a}(1,i)$\\
\STATE \hspace{1.6cm}$ K \gets K-1$\\
\STATE \hspace{1.1cm}$ \textbf{end while}$\\
\STATE \hspace{1.1cm}$ step \gets step+1$\\
\STATE \hspace{0.6cm}$ \textbf{end while}$\\
\STATE \hspace{0.2cm}$ \textbf{end for}$\\
\STATE \hspace{0.2cm}$ {{\bf{H}}_A} \gets {{\bf{H}}_a} $
\end{algorithmic}
\label{alg1}
\end{algorithm}

In the scenario where users' motion range changes, UEs from the training dataset can randomly appear in a smaller area compared with those from the testing dataset. The indoor area widens the square from 10m for the training dataset to 40m for the testing dataset and the outdoor area from 200m to 800m. All the left settings remain the same as the default setting in\cite{liu2012cost}.

The uniform linear array (ULA) model is utilized with ${N_t} = 32$ transmitting antennas and ${N_c} = 1024$ subcarriers. We also take ${N_a} = 32$ to truncate ${{\bf{H}}_a}$. The number of samples in the training dataset is 50,000 and turns to 100,000 after offline augmentation. The testing dataset includes 10,000 samples. Besides, the epochs, learning rate, batch size, and other training settings follow \cite{lu2020multi} for a fair comparison. The evaluation of performance is the normalized mean square error (NMSE) between ${{\bf{H}}_a}$ and ${{\hat{\bf{H}}}_a}$ as shown below:
\begin{equation}
{\rm{NMSE}} = E\left\{ {\frac{{\Vert{{\bf{H}}_a} - {{\hat{\bf{H}}}_a}||_2^2}}{{||{{\bf{H}}_a}||_2^2}}} \right\}.
\end{equation}

\begin{figure}[!t]
\centering
\includegraphics[width=\linewidth]{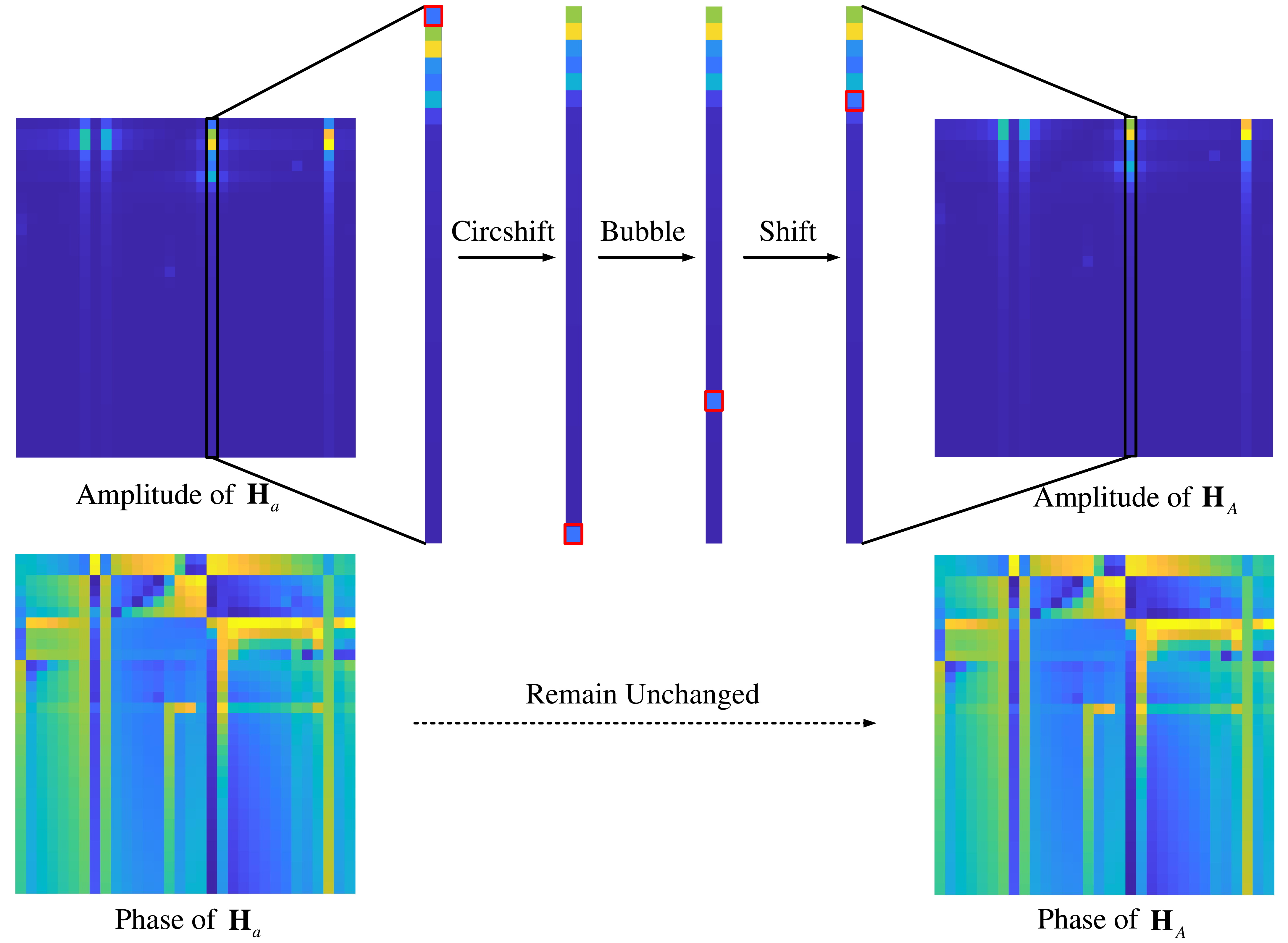}
\caption{The procedure of B-S (Upward) with $S=1$. Each column of the amplitude needs to be circularly shifted and each element can bubble to find its proper location. The phase of $\bf{H}_a$ remains unchanged.}
\label{fig_2}
\end{figure}

\begin{figure}[!t]
\centering
\includegraphics[width=\linewidth]{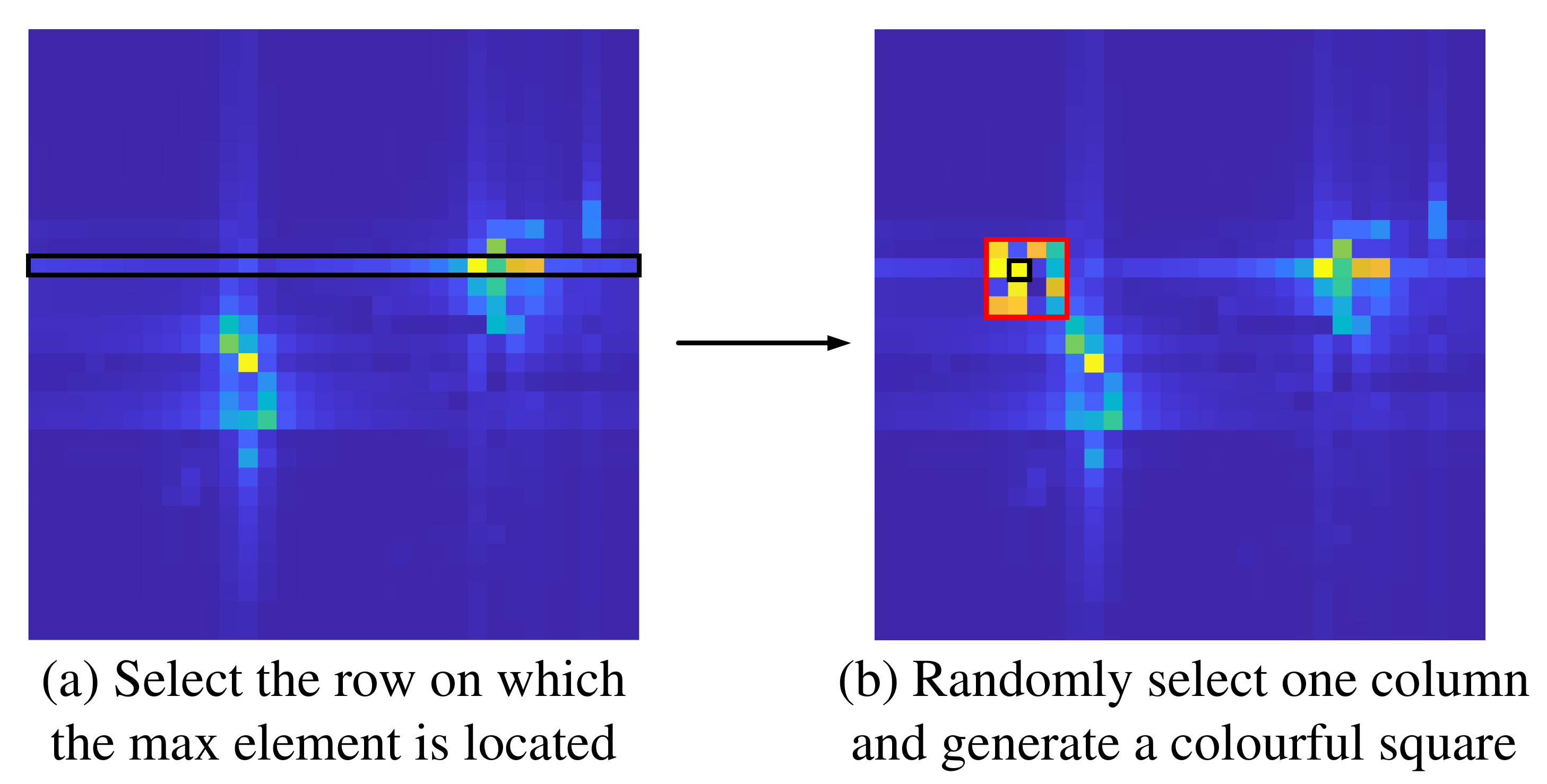}
\caption{The schematic of R-G with $k=4$.}
\label{fig_3}
\end{figure}
\subsection{Performance of The Proposed B-S and R-G}
We introduce the model-driven augmentation approach in \cite{liu2022training} which also uses the cyclic shifting as a baseline for reasonable comparison. In the scenario where users' motion mode changes, we use B-S (Upward)  and R-G algorithms to enhance the training dataset. $S$ is set to 2 in B-S (Upward) and $k$ is set to 4 in R-G data augmentation. All the augmentation approaches are tested with the CRNet structure to have a fair comparison. The performances are listed in Table I, where ``No Aug'', ``M-D'' and ``B-S'' indicate CRNet with no data enhancement, the model-driven augmentation approach in \cite{liu2022training} ($S=2$ is also set.) and our proposed B-S, respectively. ``B-S\&R-G'' represents that both B-S and R-G augmentation are employed with CRNet. 

As shown in Table I, CRNet without any augmentation suffers from a decline in accuracy compared with results in \cite{lu2020multi}, where the training dataset and the testing dataset share the same distribution in the motion mode. That means there exists a domain gap between the training dataset and the testing dataset due to the different motion modes. Our proposed B-S (Upward) has greatly improved the accuracy in the indoor environment, which indicates that the domain gap is caused by the time delay as our assumption. B-S (Upward) shortens the time delay in the training dataset, narrowing the delay gap.

\begin{figure}[!t]
\centering
\includegraphics[width=\linewidth]{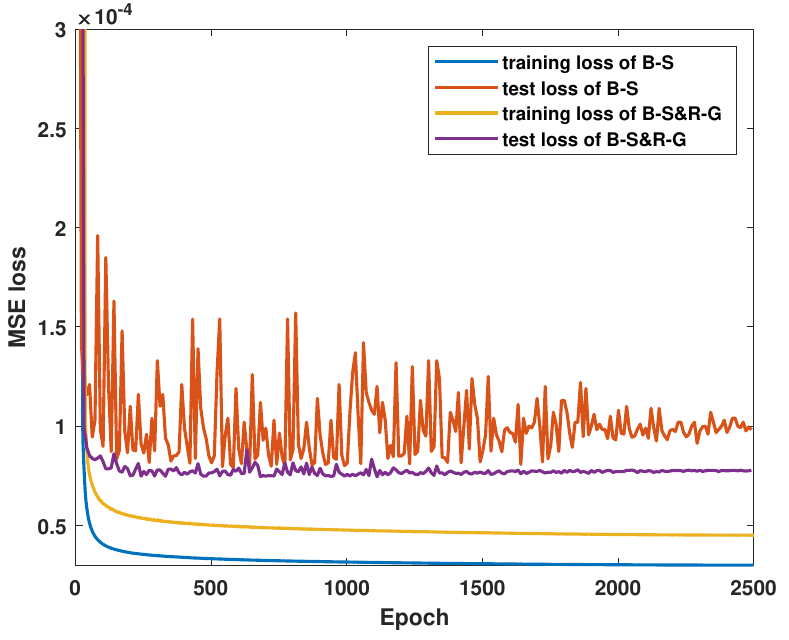}
\caption{MSE loss descending trends comparison between B-S and B-S\&R-G.}
\label{fig_4}
\end{figure}

Though ``M-D'' gets close to ``B-S'' at the outdoor scene of 1/4 and 1/32 compression rate, the R-G algorithm highlights the channel information and ``B-S\&R-G'' is the most superior. Although ``B-S\&R-G'' has higher MSE loss in the training procedure, it achieves a lower loss under test than ``B-S'' shown in Fig. 4. With the help of R-G, CRNet learns more generalized features and thus the overfitting problem is suppressed. 

The impact of hyperparameter $k$ in R-G is also tested, which is demonstrated in Table III. When $k=3$, the generated area is small and the enhancement effect of CSI features is not obvious. If we excessively enlarge the random generation region setting $k=6$, the picture of the CSI matrix can be more complex and irregular. Excessively big or small settings of $k$ both cause performance deterioration. 

\begin{table*}[htbp]
\caption{NMSE(dB) of CRNet with Existing and Our Approaches in Motion Mode Changing Scenario \label{tab:table1}}
\centering
\begin{tabular}{c|cc|cc|cc|cc|cc}
\hline
\multicolumn{1}{c|}{$\eta$}                                 & \multicolumn{2}{c|}{1/4}                                                    & \multicolumn{2}{c|}{1/8}                                                    & \multicolumn{2}{c|}{1/16}                                                   & \multicolumn{2}{c|}{1/32}                                                   & \multicolumn{2}{c}{1/64}                                                   \\ \hline
\multicolumn{1}{c|}{\multirow{2}{*}{\textbf{Methods}}} & \multicolumn{2}{c|}{NMSE}                                                   & \multicolumn{2}{c|}{NMSE}                                                   & \multicolumn{2}{c|}{NMSE}                                                   & \multicolumn{2}{c|}{NMSE}                                                   & \multicolumn{2}{c}{NMSE}                                                   \\
\multicolumn{1}{c|}{}                                  & \multicolumn{1}{c}{\textbf{Indoor}} & \multicolumn{1}{c|}{\textbf{Outdoor}} & \multicolumn{1}{c}{\textbf{Indoor}} & \multicolumn{1}{c|}{\textbf{Outdoor}} & \multicolumn{1}{c}{\textbf{Indoor}} & \multicolumn{1}{c|}{\textbf{Outdoor}} & \multicolumn{1}{c}{\textbf{Indoor}} & \multicolumn{1}{c|}{\textbf{Outdoor}} & \multicolumn{1}{c}{\textbf{Indoor}} & \multicolumn{1}{c}{\textbf{Outdoor}} \\ \hline
No Aug                                                  & -22.26                              & -7.84                                 & -11.20                              & -6.04                                 & -9.56                               & -4.30                                 & -7.37                               & -2.39                                 & -4.55                               & -0.37                                \\
M-D \cite{liu2022training}                                              & -26.09                              & -9.48                                 & -16.05                              & -7.04                                 & -10.98                              & -5.37                                 & -8.93                               & -3.60                                 & -5.32                               & -0.75                                \\
B-S                                             & \textbf{-28.22}                     & -9.41                                 & \textbf{-17.73}                     & -7.10                                 & \textbf{-13.02}                     & -5.45                                 & \textbf{-10.55}                     & -3.43                                 & \textbf{-6.65}                      & -1.28                                \\
B-S\&R-G                                          & \textbackslash{}                    & \textbf{-9.76}                        & \textbackslash{}                    & \textbf{-7.32}                        & \textbackslash{}                    & \textbf{-5.56}                        & \textbackslash{}                    & \textbf{-3.80}                        & \textbackslash{}                    & \textbf{-1.45}                       \\ \hline
\end{tabular}
\end{table*}

\begin{table*}[htbp]
\centering
\caption{NMSE(dB) of CRNet with Existing and Our Approaches in Motion Range Changing Scenario \label{tab:table2}}
\begin{tabular}{c|cc|cc|cc|cc|cc}
\hline
\multicolumn{1}{c|}{$\eta$}                                 & \multicolumn{2}{c|}{1/4}           & \multicolumn{2}{c|}{1/8}           & \multicolumn{2}{c|}{1/16}          & \multicolumn{2}{c|}{1/32}          & \multicolumn{2}{c}{1/64}           \\ \hline
\multicolumn{1}{c|}{\multirow{2}{*}{\textbf{Methods}}} & \multicolumn{2}{c|}{NMSE}          & \multicolumn{2}{c|}{NMSE}          & \multicolumn{2}{c|}{NMSE}          & \multicolumn{2}{c|}{NMSE}          & \multicolumn{2}{c}{NMSE}           \\
\multicolumn{1}{c|}{}                                  & \textbf{Indoor} & \textbf{Outdoor} & \textbf{Indoor} & \textbf{Outdoor} & \textbf{Indoor} & \textbf{Outdoor} & \textbf{Indoor} & \textbf{Outdoor} & \textbf{Indoor} & \textbf{Outdoor} \\ \hline
No Aug                                                  & -19.88          & -5.66            & -10.14          & -3.25            & -6.01           & -1.62            & -4.14           & -0.87            & -2.34           & -0.57            \\
M-D \cite{liu2022training}                                            & -22.03          & -6.21            & -10.90          & -4.46            & -8.69           & -1.84            & -5.92           & -1.16            & -4.58           & -0.65            \\
B-S                                          & \textbf{-26.97} & \textbf{-7.29}   & \textbf{-16.10} & \textbf{-4.92}   & \textbf{-9.52} & \textbf{-2.01}   & \textbf{-8.37}  & \textbf{-1.20}   & \textbf{-5.03}  & \textbf{-0.69}   \\ \hline
\end{tabular}
\end{table*}
In the scenario where users’ motion range changes, B-S (Downward) is used because UEs in the testing dataset are further away from the BS and the CSI matrix may have a longer time delay. It is hard for the network to process these samples with a longer time delay. 

We set $S=1$ for B-S (Downward) and ``M-D''. Table II shows that ``B-S'' outperforms ``M-D'' in all the compression rates under both indoor and outdoor scenarios. It is worth noting that the NMSE of CRNet is -5.66dB with $\eta=1/4$ outdoor, meaning that this scenario is exceedingly bad and the features of the CSI matrix are quite uncertain. Therefore, R-G is not added to the network since it may result in more confusion about matrix features and lose the enhancement ability.

The step of cyclic shifting has a significant influence on the result. We demonstrate the decline of performance as $S$ increases in the indoor scenario of users' motion range changing in Table IV. Since the general delay in the indoor environment is relatively short, elements of huge values are mostly distributed on the first few rows of the matrix. As a result, we must ensure $S$ in a limited degree avoiding excessive increase of time delay. 

\section{Conclusion}
In this letter, we proposed the data enhancement approaches for DL-based massive MIMO CSI feedback task. We introduced B-S algorithm making up the time domain gap caused by diverse motion modes and ranges between the training dataset and the real-time dataset (testing dataset) in practical use. Besides, R-G data augmentation was designed to enhance features in the outdoor channel since the outdoor scene was more complicated. Experiments showed that our augmentation improved network robustness.


%

\begin{table}[]
\centering
\caption{NMSE(dB) Comparison of B-S\&R-G with Different $k$ Setting in Motion Mode Changing Outdoor Scenario\label{tab:table3}}
\begin{tabular}{c|c|c|c|c|c}
\hline
\diagbox{$k$}{$\eta$} & 1/4   & 1/8   & 1/16  & 1/32  & 1/64  \\ \hline
3 & -9.42 & -7.33 & -5.38 & -3.79 & -1.41 \\
4 & \textbf{-9.76} & -7.32 & -5.56 & \textbf{-3.80} & \textbf{-1.45} \\
5 & -9.52 & \textbf{-7.46} & \textbf{-5.80} & -3.68 & -1.18 \\
6 & -9.36 & -7.10 & -5.14 & -2.88 & -0.70    \\ \hline
\end{tabular}
\end{table}

\begin{table}[]
\centering
\caption{NMSE(dB) Comparison of B-S with Different $S$ Setting in Motion Range Changing Indoor Scenario\label{tab:table4}}
\begin{tabular}{c|c|c|c|c|c}
\hline
 \diagbox{$S$}{$\eta$} & 1/4             & 1/8             & 1/16            & 1/32           & 1/64           \\ \hline
0 & -19.88          & -10.14           & -6.01           & -4.14          & -2.34          \\
1 & \textbf{-26.97} & \textbf{-16.10} & \textbf{-9.52} & \textbf{-8.37} & \textbf{-5.03} \\
2 & -22.09          & -8.79           & -5.35           & -3.69          & -2.44          \\
3 & -18.64          & -7.69           & -3.39           & -2.85          & -1.65          \\ \hline
\end{tabular}
\end{table}
\bibliographystyle{IEEEtran}
\bibliography{IEEEabrv,ref}

\begin{thebibliography}{10}
\providecommand{\url}[1]{#1}
\csname url@samestyle\endcsname
\providecommand{\newblock}{\relax}
\providecommand{\bibinfo}[2]{#2}
\providecommand{\BIBentrySTDinterwordspacing}{\spaceskip=0pt\relax}
\providecommand{\BIBentryALTinterwordstretchfactor}{4}
\providecommand{\BIBentryALTinterwordspacing}{\spaceskip=\fontdimen2\font plus
\BIBentryALTinterwordstretchfactor\fontdimen3\font minus
  \fontdimen4\font\relax}
\providecommand{\BIBforeignlanguage}[2]{{%
\expandafter\ifx\csname l@#1\endcsname\relax
\typeout{** WARNING: IEEEtran.bst: No hyphenation pattern has been}%
\typeout{** loaded for the language `#1'. Using the pattern for}%
\typeout{** the default language instead.}%
\else
\language=\csname l@#1\endcsname
\fi
#2}}
\providecommand{\BIBdecl}{\relax}
\BIBdecl

\bibitem{chen2016measurement}
J.~Chen, X.~Yin, and S.~Wang, ``Measurement-based massive mimo channel modeling
  in 13--17 ghz for indoor hall scenarios,'' in \emph{IEEE International
  Conference on Communications (ICC)}, 2016, pp. 1--5.

\bibitem{vieira2014flexible}
J.~Vieira, S.~Malkowsky, K.~Nieman, Z.~Miers, N.~Kundargi, L.~Liu, I.~Wong,
  V.~{\"O}wall, O.~Edfors, and F.~Tufvesson, ``A flexible 100-antenna testbed
  for massive mimo,'' in \emph{IEEE Globecom Workshops (GC Wkshps)}, 2014, pp.
  287--293.

\bibitem{9695409}
X.~Shi, J.~Wang, and J.~Song, ``Triple-structured sparsity-based channel
  feedback for ris-assisted mu-mimo system,'' \emph{IEEE Communications
  Letters}, vol.~26, no.~5, pp. 1141--1145, 2022.

\bibitem{wen2018deep}
C.-K. Wen, W.-T. Shih, and S.~Jin, ``Deep learning for massive mimo csi
  feedback,'' \emph{IEEE Wireless Communications Letters}, vol.~7, no.~5, pp.
  748--751, 2018.

\bibitem{lu2020multi}
Z.~Lu, J.~Wang, and J.~Song, ``Multi-resolution csi feedback with deep learning
  in massive mimo system,'' in \emph{IEEE International Conference on
  Communications (ICC)}, 2020, pp. 1--6.

\bibitem{ji2021clnet}
S.~Ji and M.~Li, ``Clnet: Complex input lightweight neural network designed for
  massive mimo csi feedback,'' \emph{IEEE Wireless Communications Letters},
  vol.~10, no.~10, pp. 2318--2322, 2021.

\bibitem{liu2022training}
Z.~Liu and Z.~Ding, ``Training deep learning models for massive mimo csi
  feedback with small datasets in new environments,'' \emph{arXiv preprint
  arXiv:2211.14785}, 2022.

\bibitem{ji2022enhancing}
S.~Ji and M.~Li, ``Enhancing deep learning performance of massive mimo csi
  feedback,'' \emph{arXiv preprint arXiv:2208.11333}, 2022.

\bibitem{xiao2022ai}
H.~Xiao, Z.~Wang, D.~Li, W.~Tian, X.~Liu, W.~Liu, S.~Jin, J.~Shen, Z.~Zhang,
  and N.~Yang, ``Ai enlightens wireless communication: A transformer backbone
  for csi feedback,'' \emph{arXiv preprint arXiv:2206.07949}, 2022.

\bibitem{hinton2006reducing}
G.~E. Hinton and R.~R. Salakhutdinov, ``Reducing the dimensionality of data
  with neural networks,'' \emph{science}, vol. 313, no. 5786, pp. 504--507,
  2006.

\bibitem{liu2012cost}
L.~Liu, C.~Oestges, J.~Poutanen, K.~Haneda, P.~Vainikainen, F.~Quitin,
  F.~Tufvesson, and P.~De~Doncker, ``The cost 2100 mimo channel model,''
  \emph{IEEE Wireless Communications}, vol.~19, no.~6, pp. 92--99, 2012.

\end{thebibliography}



\end{document}